\documentclass[prl,twocolumn,showpacs,superscriptaddress]{revtex4}% Physical Review B
\usepackage[dvips]{graphicx}% Include figure files
\usepackage{dcolumn}% Align table columns on decimal point
\usepackage{bm}% bold math
\usepackage{amssymb}
\usepackage{textcomp}

\DeclareGraphicsExtensions{eps}

\begin{document}
\preprint{APS/123-QED}
\title{$^{17}$O NMR study of the intrinsic magnetic susceptibility and spin dynamics of the quantum kagom\'{e} antiferromagnet ZnCu$_3$(OH)$_6$Cl$_2$}% Force line breaks with \\
\author{A. Olariu}
\author{P. Mendels}
\author{F. Bert}
\affiliation{%
Laboratoire de Physique des Solides, Universit\'{e}
Paris-Sud, UMR CNRS 8502,  91405 Orsay, France}%
\author {F. Duc}
\affiliation{%
Centre d'\'Elaboration des Mat\'eriaux et d'\'Etudes
Structurales, CNRS UPR 8011, 31055 Toulouse, France}%
\author {J.C. Trombe}
\affiliation{%
Centre d'\'Elaboration des Mat\'eriaux et d'\'Etudes
Structurales, CNRS UPR 8011, 31055 Toulouse, France}%
\author{M.A. de Vries}
\affiliation{%
School of Chemistry and CSEC, The University of Edinburgh, Edinburgh, EH9 3JZ, UK  }%
\author{A. Harrison}
\affiliation{%
School of Chemistry and CSEC, The University of Edinburgh, Edinburgh, EH9 3JZ, UK  }%
\affiliation{ Institut Laue Langevin, F38042 Grenoble, France}
\date{\today}% It is always \today, today,

\begin{abstract}
We report through $^{17}$O NMR, an unambiguous local determination of the intrinsic kagom\'{e} lattice spin susceptibility as well as that created around non-magnetic defects arising from natural Zn/ Cu exchange in the $S$=1/2 (Cu$^{2+}$)
herbertsmithite ZnCu$_{3}$(OH)$_6$Cl$_2$ compound. The issue of a singlet~-~triplet gap is addressed.  The magnetic response around a defect is found to markedly differ from that observed in non-frustrated antiferromagnets. Finally, we discuss our relaxation measurements  in the light of Cu and Cl NMR data and suggest a flat q-dependence of the excitations.
\end{abstract}

%\pacs{75.40.Gb, 76.75.+i}% PACS, the Physics and Astronomy
\maketitle

When nearest neighbor interacting 1/2 spins decorate the vertices of
corner sharing triangles, as in the kagom\'e antiferromagnet (AF),
the geometry of the lattice efficiently frustrates the magnetic
interactions and the classical N\'eel states are certainly
destabilized. Since Anderson's seminal idea of a resonating valence
bond spin liquid state~\cite{Anderson}, the actual ground state has
been a matter of deep conceptual debates and various theoretical
proposals were made in the last 15
years~\cite{Lecheminant,Sachdev,Hastings,Mila,Misguich,Nikolic,Budnik,Wang,Ran}
with no definitive conclusion so far. The experimental situation is
quite contrasted and has been lagging well behind theory for long,
unable to clarify the pending debates on the nature of the kagom\'e
antiferromagnet ground state. The main reason is that achieving well
decoupled perfect kagom\'{e} planes has represented a real challenge
since the pioneering work on kagom\'{e} bilayers in
SrCr$_8$Ga$_4$O$_{19}$~\cite{SCGORamirez}.

Quite recently the end member ($x=1$) of the paratacamite family
Zn$_x$Cu$_{4-x}$(OH)$_6$Cl$_{2}$, known as herbertsmithite, under
the name of the discovered mineral, was revealed as a "structurally
perfect $S$=1/2 kagom\'e antiferromagnet"~\cite{Shores:05} and
opened a new avenue to comparisons between experiment and theory.
Due to a more favorable electrostatic environment, Cu$^{2+}$ is
expected to preferentially occupy the distorted octahedral kagom\'e
sites. Therefore ZnCu$_3$(OH)$_6$Cl$_2$ could feature perfect
Cu$^{2+}$, $S$=1/2, kagom\'e planes separated by non magnetic
Zn$^{2+}$ layers (Fig.~1). The lack of observation of any order down
to the lowest explored temperature (50
mK$\sim$~$J$/3000~-~J is the exchange interaction)~\cite{Mendels,Helton,Ofer} is a clear indication
that this material is the best to-date to approach the ideal physics
of a kagom\'e lattice.

Since the very first studies, a debate has settled about the actual
low-$T$ susceptibility $\chi$, one of the most crucial tests of the
low-$T$ ground state. Would a singlet ground state occur, $\chi$
should yield the singlet-triplet gap. In complete contradiction with
this picture, the macroscopic susceptibility $\chi_{macro}$, as
measured by SQUID, shows a monotonous increase down to $T$
=0~\cite{Helton}. Although additional Dzyaloshinski-Moriya (DM)
interactions were proposed to explain such a low-$T$
increase~\cite{Rigol}, recent neutron structural refined
data~\cite{Lee,deVries} rather indicate 5-10\% Zn$^{2+}$/Cu$^{2+}$
inter-site exchange which could be responsible for a Schottky-like
field dependent anomaly detected in specific heat~\cite{deVries} and
be the source of all or part of the low-$T$ increase of $\chi_{macro}~$
~\cite{Bert}. This would result from both
isolated Cu$^{2+}$ spins on the Zn$^{2+}$ site which are weakly
coupled to the kagom\'e layers and non-magnetic Zn$^{2+}$ defects in
the Cu$^{2+}$ kagom\'e planes which are known to generate a specific
spin texture in their vicinity as was suggested in former studies of
magnetic dilution of frustrated lattices~\cite{MendelsSCGO,
Dommange, Bert05proceeding}.

Local probes, such as used in nuclear magnetic resonance (NMR), are
invaluable to sort out the various contributions to $\chi$. They
enable one to map out a histogram of susceptibilities at various
sites, otherwise summed up in $\chi_{macro}$. $^{17}$O NMR of
enriched samples appears as the best compromise for such a local
study. Due to the strong magnetic Cu-Cu bond through the bridging
oxygen, one can indeed safely expect that O dominantly couples to
the 2 neighboring Cu sites of the kagom\'e planes only (Fig.~3)
while its detection is not wiped-out through relaxation effects at
variance with the case of Cu~\cite{Imai}. For the first
time, we are able to track accurately the kagom\'e lattice
susceptibility from 300 to 0.47~K, and also, beyond establishing
definitely the existence of Zn/Cu substitution, use the latter
perturbation to reveal some original facets of the kagom\'e physics
around these defects. We also present a comparison of our $^{17}$O
NMR relaxation data with that from Cl and Cu sites
\cite{Imai} , which suggests a non-dispersive branch of low energy
excitations.

The NMR experiments were performed on several unaligned powder
samples of different origins which gave identical results.
Preparation and sample characterization follow standard routes
described elsewhere \cite{Shores:05,Mendels} except that 50\%
$^{17}$O enriched water was used for the preparation and samples
were not washed after filtration. $^{17}$O NMR spectra were recorded
from room-$T$ down to 0.47~K by integrating the echo recorded by
sweeping the field at a fixed frequency of 38.974~MHz. Delays
between pulses were 10-20~$\mu$s, except in contrast experiments
performed at low-$T$ where delays, up to 200~$\mu$s allow to single
out slowly relaxing sites. Quadrupolar and magnetic contributions
were separated by taking additional spectra at 19.974 and 29.974 MHz. Quadrupolar singularities typical of a powder pattern
are easily observed on the wings of our spectra at high-$T$ (Fig.~1) and both a
quadrupolar frequency  $\nu_Q$~=~0.70(3)~MHz and the asymmetry
parameter $\eta= 0.64(10)$ were determined by the extreme cut-offs
in the powder pattern ($^{17}$I = 5/2).
%A spurious contribution
%(cross on Fig.~1) which disappears below 150~K, was also observed at
%high $T$.
The $T_1$ recovery curves were obtained by irradiating the
main line with a train of 5 pulses separated by $\sim T_2$. Each
component mixes all nuclear transitions (powder pattern) and we
fixed the coefficients of the recovery law at the $T$~=~70~K values.
%Due to the proximity of the (D) line, a slow relaxing contribution was found at low T. We therefore allowed a second component in our fits.

In the present letter, we track the evolution of the central part of
the NMR spectrum. In the ideal case of a fully occupied Cu$^{2+}$
kagom\'e plane, the structure yields a single environment of the O
site therefore one should observe a unique central $-1/2 \rightarrow
1/2$ sharp singularity in the high-$T$ range ($T > J$), the shift of
which reflects the susceptibility of the two neighboring Cu.  On the
contrary, two different lines rather than one, corresponding to
central transitions are surprisingly evident in our $T =300$~K
spectrum [(M) and (D) in Fig.~1].
  \begin{figure}
\includegraphics[width=7cm]{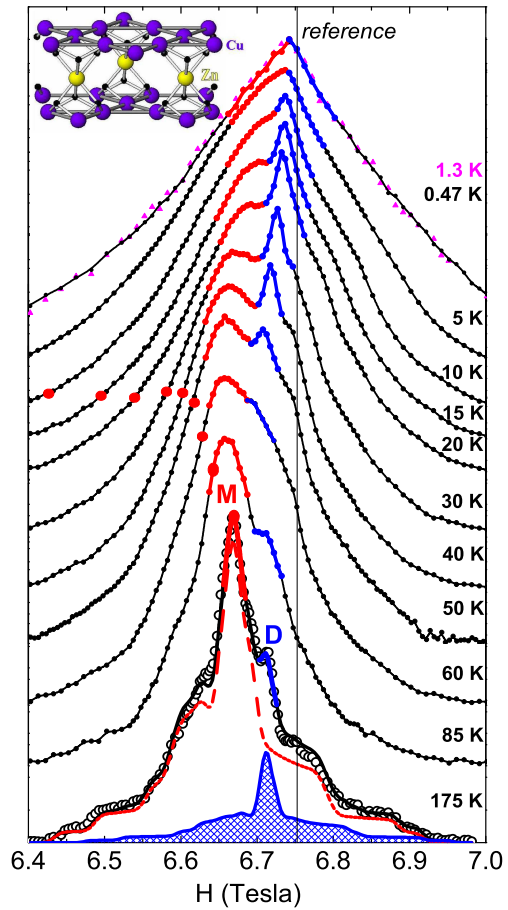}
\caption{$^{17}$O NMR spectra from 300~K to 0.47~K. The shift
reference is indicated by the vertical line.
%The $T$-dependent shift
%measured from this reference tracks the intrinsic susceptibility of
%the kagom\'e planes and its distribution. The cross on the 300~K
%spectrum relates to a spurious phase no more detected at 150~K.
The line on the 175~K spectrum is a simulation using two components,
 (M, dotted line) and (D, hatched), see text.
Large full circles indicate the position of the center of the line
which would be expected from $\chi_{macro}$.}
\end{figure}
This clearly indicates the existence of \emph{two} distinct local
magnetic environments. The ratio of the magnetic shifts $\sim 2$ for
the two lines (M) and (D) naturally supports the association of the
most shifted and most intense line, with the vast majority of O
nuclei locally coupled to two magnetic Cu, while for the
other line~(D), O couples to only one Cu, hence the other
neighboring site is occupied by Zn. Since for each Zn at a Cu site
of the kagom\'e plane four O are affected, the 20(5)\% intensity of
this line is found consistent with a 6(2)\% Zn/Cu substitution
suggested by other techniques~\cite{Helton,Lee,deVries,Bert}.

The shift of both lines, $K^{(D)}$ and $K^{(M)}$ can be easily
tracked in a broad $T$-range. The main line~(M) gives access to the
intrinsic susceptibility of the kagom\'e lattice through the line
position while the physics of a Cu 1st near neighbor of a
non-magnetic defect is best probed through the other line (D). The
positions of the (M) and (D) lines were found to scale with the
field after correction of second order quadrupolar effects for low
frequencies, which proves their magnetic origin. No simulation even
allowing different axis for the EFG and shift tensors could account
for the presence of these two peaks in a single O site model.
\begin{figure}
\includegraphics[width=7cm]{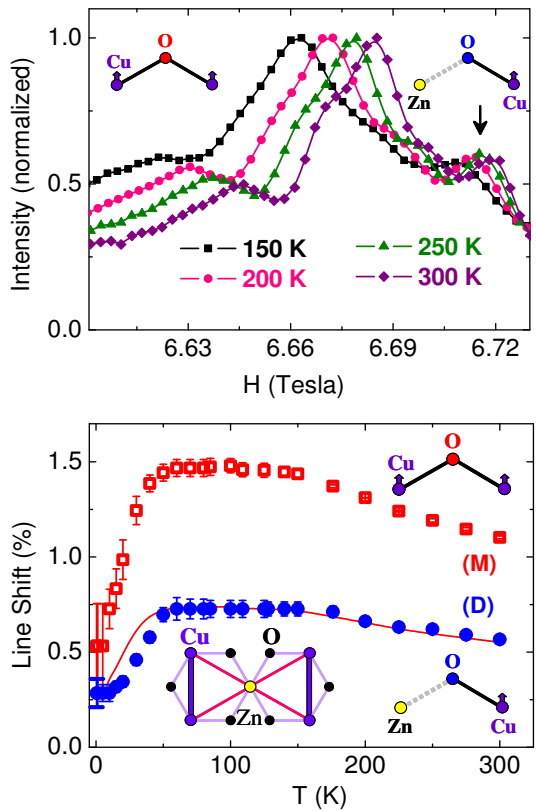}
\caption{Top: blow-up of high $T$ spectra. The shift of
the main and defect lines are evident and found different. Bottom:
Plot of the magnetic shift of both lines versus $T$. The line
through the $K^{(D)}$ plot represents $K^{(M)}$/2. The sketch on
lower left corner is for what would be expected around a Zn
substituted on the Cu kagom\'e plane at $T=0$ and thick lines
represent Cu-Cu dimers.}
\end{figure}

The (M) line is found to shift towards low fields (increase of
susceptibility) between 300 and 100~K (Fig.~2a), goes through a
maximum between 100 and 50~K and then progressively moves back
towards the reference below 40~K while a rapid increase of the width
occurs (Fig.~1). In the low-$T$ regime, below 5~K, the linewidth
tends to saturate and no difference was found between spectra taken
at 1.3 and 0.47~K. The variation of the shift, $K^{(M)}$, is
reported in Fig.~2b. It monitors the $T$-evolution of the
susceptibility of Cu in the kagom\'e planes, except those sitting
nearby non-magnetic defects, through $K^{(M)}=2A \chi_{{\rm Cu}}$,
with the hyperfine constant $A$~=~35(2)~kOe/$\mu_{{\rm
B}}$~\cite{shift}. It is worth noting that this coupling value is
approximately 30 times larger than those reported for Cl NMR and
$\mu$SR experiments, hence the invaluable sensitivity
of the $^{17}$O probe.

Two main model-free results can be deduced from $K^{(M)}$: (i) A
maximum of the kagome susceptibility is found to occur around $J/2$
which differs markedly from $\chi_{macro}$. Defects induced by Cu/Zn
substitution and responsible for the low $T$ increase of
$\chi_{macro}$, contribute to the NMR linewidth. (ii) The
susceptibility reaches a non-zero value at low $T$, typically 1/3 of
the maximum value which indicates the absence of a gap in
herbertsmithite.

%We now address the central issue of the existence of a gap and compare our data with numerical calculations.
One can
invoke three possible scenarii: (i) a \emph{non-singlet} ground
state of the ideal kagom\'e system, such as observed here, with a
decreasing susceptibility at low $T$ associated with the
strengthening of the AF correlations below $T \simeq J/2$ which then
level off as $T \rightarrow 0$, and surprisingly, do not yield a
gapped ground state. (ii) Sticking to a \emph{singlet-triplet gap}:
The finite $T$ = 0 susceptibility observed in herbertsmithite may
result from a filling of the hypothetical gap -expected to be at
most $J/20$- under the modest applied field 6.5~T of our experiment.
It is however not clear to us whether the magnetic energy brought to
the system $\sim J$/30 should be compared to the gap value or to the
larger one at which a magnetization plateau should occur. In
addition, the value of the maximum of $K^{(M)}$ around 50~K is not
on-line with the expected small gap value. (iii) The existence of
additional DM interaction
$\overrightarrow{D}.(\overrightarrow{S}_i\times
\overrightarrow{S}_j$) between Cu$^{2+}$ spins which results from
the absence of inversion center between Cu, as earlier
suggested~\cite{Rigol}. This would modify the $T$=0 susceptibility
by \emph{mixing singlet and triplet} states. Although  $\chi$
could not be calculated below 0.25 $J$~\cite{Rigol},
we anticipate from a comparison of our $K^{(M)}$ curve and
~\cite{Rigol} that this DM term might be smaller than suggested and
rather less than 0.05 $J$ at least for one component of
$\overrightarrow{D}$. Finally, one should point that exact
calculations, reliable down to 0.2-0.3~J, do not indicate any
maximum of $\chi$. The deviance of our experimental data from the
ideal case could be attributed either to the DM interaction or to
the rather strong dilution of the network favoring a dimerization of
the ground state (see below).

We now turn to the defect line (D) which reflects the susceptibility
of a Cu in the immediate vicinity of a non-magnetic Zn substituted
on a kagom\'e site. At high-$T$,  the (D) line follows a trend
similar to the (M) line, as can be observed through the comparison
of spectra taken between 300 and 150 K (Fig.~2a). The 1:2 scaling of
$K^{(D)}$ with $K^{(M)}$ (Fig.~2b) indicates that in this $T$-range
defects have negligible effects on the Cu sites in their immediate
vicinity, which underlines the weakness of the correlation length
even for $T\sim J$, another well-known landmark of frustration. The
(D) line surprisingly appears below 50~K as a sharp prominent
feature of the overall spectrum, which contrasts with the broadening
of the (M) line in the same $T$-range (Fig.~1). This relative
narrowness points at a reduced distribution of susceptibilities
which witnesses a quite unique environment around the defect. In
addition, below 50~K, the decrease of the shift is found sharper
than for the (M) line, an indication that the susceptibility of Cu
in the neighborhood of a Zn decreases faster. Finally, the
transverse relaxation was found substantially longer than for the
(M) line in this $T$-range which enabled us to perform contrast
experiments and better isolate the defect line.  A relative
intensity of the (D) line $\sim$ 30(4)\% can then be calculated.

The picture which can be sketched differs markedly from the
situation met in non-frustrated AF where the spin response is
commonly enhanced in the immediate vicinity of  a non-magnetic
defect~\cite{RevModPhys} and would yield a rapid low-T increase of
$K^{(D)}$ rather than a decrease as clearly observed here. Our
findings are in qualitative agreement with the results from exact
diagonalizations where a defect is found to induce a quite reduced
susceptibility or even singlets on the 4 neighboring sites while an
enhanced staggered susceptibility is induced for further
neighbors~\cite{Dommange}. The latter would be naturally responsible
for the smoothing out of the quadrupolar singularities and could
explain part or all of the broadening of the (M) line. Another
source of broadening could also result from the alternating DM
contribution as proposed recently~\cite{Rigol}. In the ideal case of
singlets induced around a non-magnetic defect, a simple counting
argument yields 6 O sites per defect (fig.~3) with a zero shift,
therefore 36\% of the O sites, in agreement with the detected
intensity. The decrease of $K^{(D)}$, sharper than $K^{(M)}$, could
therefore be explained by a tendency for Cu to dimerize near a
defect, as expected from the local relief of frustration. The fact
that we do not observe a zero-susceptibility and a gap resulting
from dimers could certainly be ascribed again to DM interactions.
\begin{figure}
\includegraphics[width=8cm]{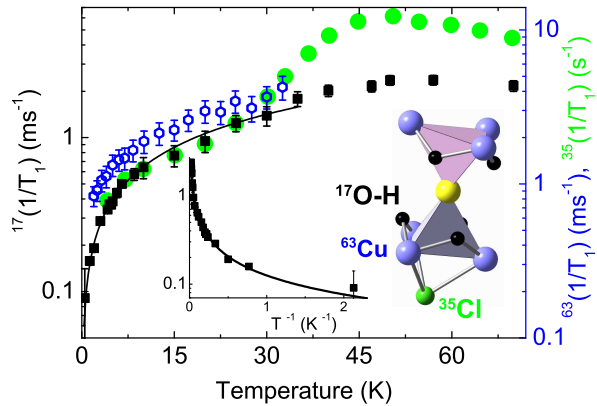}
\caption{ Semi-logarithmic $T_1^{-1}$ plots for $^{17}$O (this work)
 and for $^{63}$Cu and $^{35}$Cl~\cite{Imai}). The similarity of the
$T$-variation for the 3 nuclei is evident. Inset: $^{17}T_1^{-1}$ versus
$T^{-1}$. A non-zero gap would yield a linear variation, not
observed here.}
\end{figure}

Finally, $T_1$ measurements performed at the (M) peak yield
information about the excitations. A second contribution, likely due
to the proximity of the (D) line was found below 50~K, with a $T_1$
which gets longer when $T$ is lowered. This certainly corroborates
the decrease of magnetic effects in the vicinity of a Zn defect, as
already suggested by our shift data. In Fig.~4, we plot the short
contribution below 70~K, dominated by the physics of the kagom\'e
plane, and compare it to existing Cl and O data~\cite{Imai}. Between
70 and 30~K, we do not find a marked maximum such as observed for Cl
NMR which had been suggested to be of structural origin. Due to its
weak quadrupolar moment and its larger coupling constant to Cu,
$^{17}$O NMR is less sensitive to structural effects and probes
dominantly magnetic properties. Whatever the origin of the
$^{35}(1/T_1)$ peak, our data indicates that magnetic properties are
not affected. Like for Cu $T_1^{-1}$ , we find a downward curvature of our data
below 20~K and the inset of Fig.~3 clearly shows that the
excitations are \emph{not gapped}. The absence of anomaly in the
$T_1$ variation is on-line with the absence of signature of a
transition as observed in our shift measurements down to 0.47~K or
in earlier measurements~\cite{Mendels,Helton,Ofer}. We rather find
that $T_1^{-1} \sim T^{0.73(5)}$ and the perfect agreement of all
data taken on various probes with different local $q$-space
filtering of fluctuations is  a clear experimental indication that
the excitation modes responsible for this relaxation are
dispersionless as suggested for long~\cite{dispersionless}. One
would indeed expect that "triangular 120$^{\circ}$" modes or ($\pi$
,$\pi$) modes are filtered out and do not contribute to the
relaxation respectively at the Cl or O sites at variance with Cu.
%Therefore, two very original features of excitations emerge from these relaxation studies: the absence of a gap and a flat $q$-dependence of $\chi^{''}(q,\omega)$.

To summarize, we demonstrate the existence of a maximum in the
susceptibility at $\sim J/2$ certainly related to the kagom\'e
physics and also observed in other corner-sharing kagom\'e based
lattices~\cite{MendelsSCGO,BonoBSZNMR,Bert}. A rapid decrease of the
susceptibility to a finite value at low-$T$ and ungapped excitations
point at a non-singlet ground state in Herbertsmithite. The
excitations have not a marked structure in $q$-space, a landmark of
frustration. A depressed susceptibility is observed for Cu next to
defects. This response to defects is certainly specific to the
kagom\'e physics and observed for the first time here. It opens the
route for revealing intrinsic physics through magnetic perturbations
after a better control of Zn/Cu substitution is achieved. In order
to clarify whether a gap would occur in ideal kagom\'e Heisenberg
antiferromagnets, an accurate experimental determination of the DM
term is necessary and the search for a zero-defect "second
generation" compound guides future experimental work.

We acknowledge support from ANR under the project "OxyFonda". We
thank C~Lhuillier, G.~Misguich and P.~Sindzingre for discussions.


\begin{thebibliography}{19}
\bibitem{shift}The anisotropy of the  shift tensor was found to be less than 0.2\%,
 a value consistent with the dipolar contribution on the O site. The origin of the shift $K_{chem} = -0.08(6)$\%
  and the isotropic hyperfine constant were determined by plotting the high-T variation of the shift versus  $\chi_{macro}$
   (not shown here).
\end{thebibliography}
\end{document}